# Flow and Jamming of Granular Suspensions in Foams

B. Haffner[1], Y. Khidas[2] and O. Pitois[1]

[1] *Université Paris Est, Laboratoire Navier, UMR 8205 CNRS – École des Ponts ParisTech – IFSTTAR cité Descartes, 2 allée Kepler, 77420 Champs-sur-Marne, France*

[2] *Université Paris Est, Laboratoire Navier, UMR 8205 CNRS – École des Ponts ParisTech – IFSTTAR 5 bd Descartes, 77454 Marne-la-Vallée Cedex 2, France*

## Abstract:

The drainage of particulate foams is studied under conditions where the particles are not trapped individually by constrictions of the interstitial pore space. The drainage velocity decreases continuously as the particle volume fraction $\varphi_p$ increases. The suspensions jam – and therefore drainage stops – for values $\varphi_p^*$ which reveal a strong effect of the particle size. In accounting for the particular geometry of the foam, we show that $\varphi_p^*$ accounts for unusual confinement effects when the particles pack into the foam network. We model quantitatively the overall behavior of the suspension – from flow to jamming – by taking into account explicitly the divergence of its effective viscosity at $\varphi_p^*$. Beyond the scope of drainage, the reported jamming transition is expected to have a deep significance for all aspects related to particulate foams, from aging to mechanical properties.

## 1. Introduction:

Foams are used in a lot of industrial processes: gas is mixed in many materials to improve their thermal performance or to make them lighter, which is favourable to sustainable development. The matrix of those aerated materials is often composed of a complex fluid, such as a suspension. Typical examples for such mixtures can be found in food and cosmetic industries [1], in the production of construction materials [2] and of ceramic foams [3] that are used in numerous fields of technological processes such as filtering, membranes, catalysis, … Note also that the mining industry extensively resorts to mixtures of foam and particles through the flotation process that is used to separate ores [4].

The homogeneity of a foam sample can be drastically affected by the drainage of the liquid and the simultaneous rise of the bubbles, resulting in a degradation of the quality and the properties of the final material. During the last two decades, most of the progress realized in the field of foam drainage has concerned aqueous foams, i.e. dispersions of densely packed gas bubbles in a liquid [5]. Some very recent studies have focussed on the drainage behaviour of foamy complex fluids, such as clays [6], coal fly ashes [7], colloidal suspensions [8], granular suspensions [9], emulsions [10]. Despite the results provided by these studies, a sound understanding of drainage laws in the presence of suspended particulate matter is still lacking. In order to explain the reported drainage velocities the authors have resorted to particle trapping phenomena, which can be classified into two distinct mechanisms: (i) the individual capture of particles by the foam constrictions, and (ii) the collective trapping – jamming – of the suspension.

With regard to the first mechanism (i), Louvet et al. [11] studied the capture/release transition of a single spherical particle confined within the interstitial network of foam. The authors introduced a confinement parameter $\lambda$ which compares the size of the suspended particles to that of the foam constrictions. Afterwards, $\lambda$ has been proved to control the drainage behaviour of aqueous foams containing a moderate volume fraction of density-matched spherical particles [9]. A sharp transition has been highlighted: for $\lambda < 1$ particles are free to drain with the liquid, which involves the shear of the suspension in foam interstices, for $\lambda > 1$ particles are trapped and the mobility of the interstitial phase is strongly reduced. Moreover, simple modelling has been found to describe the reported drainage behaviour as a function of $\lambda$. This study, that involved a dedicated model experimental system, has shown a promising way to progress further in the understanding of particulate foams. In this paper we follow this approach and we investigate the second trapping mechanism (ii), i.e. the jamming of the suspension within the interstitial foam network. This phenomenon is expected to be observed at a sufficiently high particle volume fraction [12]. The foam network induces confinement constraints that could influence this jamming transition. Indeed this confinement effect has been reported in studies involving small gap sizes in conventional rheometers [13-16] as well as dedicated set-up [17]. Therefore we will pay a particular attention to this issue. In order to fully uncouple the two trapping mechanisms, we consider the situation $\lambda < 1$, for which particles are not subjected to the individual capture process (i) [9], and we measure the drainage velocity as a function of both particle volume fraction and $\lambda$.

## 2. Experimental set-up:

Particulate foams samples are prepared from a precursor liquid foam which is subsequently mixed with a granular suspension (Fig. 1). The foaming solution contains 10 g/L of TetradecylTrimethyl-Ammonium Bromide (TTAB) in distilled water with 20% w/w glycerol. With such a proportion of glycerol the density of the solution is 1050 kg/m$^3$ and matches with that of polystyrene particles used in the study. The surface tension of the liquid/gas interface is 38 mN/m and shear viscosity of the bulk is $\mu_0 \simeq 1.7$ mPa.s. As we can see on Fig. 1a, bubbles are generated in a T-junction with two entries (nitrogen and foaming solution) and one exit (bubbly solution). Thanks to the flow focusing mechanism [18], small volumes of gas and liquid pass alternatively through the junction, resulting in the production of bubbles, which size is controlled by tuning the flow rates of gas and liquid. For this study the bubble diameter has been set to $D_b = 660$ µm $\pm$ 30 µm. The bubbles are continuously produced and released at the bottom of a column which is partially filled with the foaming solution (Fig. 1b). This results in the formation of foam in the column. During the production, the foam is imbibed with the same foaming solution in order to obtain stationary drainage conditions with a constant value of the gas fraction ($\phi_1$) throughout the foam column [19]. Once the column is filled, the foam is flushed towards a mixing device (which is also based on a T-junction) where the granular suspension is introduced (Fig. 1c). The suspension is prepared at a given particle volume fraction ($\varphi_2$) by mixing the foaming solution and polystyrene spherical beads (Microbeads®). The beads are quite monodisperse: $\Delta d_p/d_p \approx 5\%$ and we have used the four following diameters: $d_p = 6, 20, 30, 40$ µm. We have checked that the mixing device does not break bubbles and therefore, the bubble size in the final sample is also $D_b = 660$ µm. The outlet of the mixing device is connected to a cylindrical tube (26 mm in diameter) in which the produced particulate foam is continuously introduced (Fig. 1d). It is equipped with a piston which rate for withdrawing motion compensates exactly the volume flow rate of the injected particulate foam. Moreover, the tube is rotated (0.3 Hz) along the horizontal axis in order to compensate the effects of gravity during the filling step. We stop this step once the volume of produced particulate foam equals 60 mL, which corresponds to a foam length approximately equal to 11.5 cm. Then the foam tube is turned to the vertical and we start to measure the drainage properties of the samples. We follow the evolution of the height $h(t)$ locating the transition between the foam and the drained suspension at the bottom of the column (see Fig. 1-II). Note that the main contribution to the global error on $h(t)$ is related to the apparent thickness of the transition due to bubble size. This error is close to $\Delta h/h \approx 15\%$ excepted for the very low values of $h(t)$.

The other parameters are controlled by the relatives flow rates of the precursor foam ($q_1$) and the suspension ($q_2$). The resulting gas fraction is $\phi = q_1\phi_1/(q_1+q_2)$. For the particle fraction, rather than considering the entire particulate foam volume, we will see that it is more appropriate to define the volume fraction of particles in the interstitial phase: $\varphi_p = q_2\varphi_2/[q_1(1-\phi_1)+q_2]$. For all the samples presented in the following we have controlled the production stage in such a way that $\phi = 0.9$. As we are interested in confinement effects on the drainage of particulate foams, we refer to the confinement parameter $\lambda$ [11,9], that compares the particle size to the size $d_C$ of passage through constrictions in the interstitial network of the foam. In [11],

$$\lambda = \frac{d_p}{d_C} = \frac{1+0.57(1-\phi)^{0.27}}{0.27\sqrt{1-\phi}+3.17(1-\phi)^{2.75}}\frac{d_p}{D_b} \quad (1)$$

has been determined from both experiments involving the trapping/release of a single particle in foams and numerical simulations of foam structures. Using the values for $d_p$, $D_b$ and $\phi$, we obtain for the following $\lambda$-values probed in this study: $\lambda = 0.13, 0.43, 0.65, 0.87$.

## 3. Kinetics of drainage:

For the particle free samples, Fig. 2a shows the measured curve $h(t)/h_0^\infty$, where $h_0^\infty \equiv h(t\to\infty, \varphi_p = 0)$: a first stage is characterized by a rapid linear increase for times $t<\tau$ (inset Fig. 2a), followed by a slower evolution towards the equilibrium value $h_0^\infty$. The time $t=\tau$, is identified as the characteristic time for which half of the liquid volume has drained off the foam [5]. During this regime, the volume of liquid/suspension drained out of the foam has flowed through foam areas that have not yet been reached by the drainage front, i.e. areas where the gas fraction has remained equal to the initial value $\phi$. Because the linear regime accounts for drainage properties of foam characterized by a constant gas fraction $\phi$, we measure the drainage velocity $V$ from the slope of this linear evolution, $V = dh/dt$. In order to characterize the effect of particles on drainage, we normalize the measured drainage velocity by the one measured without particle, i.e. $V/V_0$. Note that because of uncertainties related to the measurement of $h(t)$ for $h \simeq 0$, linear fits are not applied to the early stage of the linear regime. Consequently, the relative error on the reduced drainage velocity is estimated to be close to 15%.

Fig. 2 illustrates the measured evolutions for $h(t)/h_0^\infty$ as $\varphi_p$ (Fig. 2a) or $d_p$ (Fig. 2b) varies. Both parameters modify significantly the drainage of particulate foams: (i) the initial slope decreases as both $\varphi_p$ and $d_p$ increase, (ii) the final value $h_{\varphi_p}^\infty \equiv h(t \to \infty, \varphi_p)$ decreases as well. The linear regime remains rather well defined for each sample, which suggests that the slope reasonably accounts for drainage corresponding to stationary conditions within the imposed initial conditions. $h_{\varphi_p}^\infty$ accounts for the final retention level for particles in the foam. Even if particles are not captured during the linear regime of drainage ($\lambda < 1$ within these drainage conditions), they get trapped as the drainage front reaches them and imposes the condition $\lambda > 1$. The larger the particles are, the earlier they get trapped when the drainage front goes down, and the higher the retention level is. Note that images from the bottom of the foam column confirmed that the released particles are effectively released during the first regime of drainage [9]. On Fig. 2c, reducing $h(t)$ by $h_{\varphi_p}^\infty$ and $t$ by $\tau$ make all the curves of Fig. 2a and 2b collapse onto a single one. Note that although $\tau$ and $h_{\varphi_p}^\infty$ vary significantly from one sample to the other, this confirms that free-particle and particulate foams exhibit the same drainage behavior.

All the drainage velocities are now plotted in Fig. 3. For the different values of $\lambda$, it shows a regular decrease of $V/V_0$ as the particle volume fraction $\varphi_p$ increases. The effect of particle size is not significant as $\varphi_p \lesssim 0.2$, but discrepancies appear for larger values. Drainage velocities seem to vanish, i.e. $V/V_0 \approx 0$, as $\varphi_p$ reaches approximately 0.5. The inset in Fig. 3 reveals that the particular value of $\varphi_p$ for which the drainage velocity vanishes increases with the particle size.

## 4. Discussion:

Drainage experiments have provided results for the flow of granular suspensions through the interstices of foams. As a starting point, we analyze these results in terms of the reduced effective viscosity of the suspension, i.e. $\tilde{\mu}_{eff}(\varphi_p) = \mu_{eff}(\varphi_p)/\mu_0$ , which is deduced from the drainage velocities through the relation: $\tilde{\mu}_{eff}(\varphi_p) = V_0/V(\varphi_p)$ [5]. Fig. 4 shows this quantity as a function of $\varphi_p$ for the four studied values of $\lambda$. For $\varphi_p \lesssim 0.2$ the viscosity of the suspension is consistent with the theoretical values for the bulk viscosity of diluted non-brownian solid spheres estimated with the expression $\tilde{\mu}_{eff}(\varphi_p) = 1 + 2.5\varphi_p + 6.95\varphi_p^2$ [20]. Whereas this agreement is expected for $\lambda \ll 1$, one can question the agreement observed for $\lambda \approx 1$. It should be realized that for the rather wet foams considered here, the suspension is mostly contained within the foam nodes and the volume of a foam node, $\upsilon_n$, is large enough to be a representative volume of suspension. $\upsilon_n$ can be estimated in assuming 6 nodes per bubble [5]: $\upsilon_n \approx \pi(1-\phi)D_b^3/36\phi$. In relating the bubble size to the radius of Plateau borders $r_{Pb}$ through $r_{Pb}/D_b \approx 0.62(1-\phi)^{0.45}$ [5], the node volume reads $\upsilon_n \approx 0.9r_{Pb}^3$, or equivalently $\upsilon_n \approx 30d_p^3$, which corresponds approximately to 60 sphere volumes. This means that although the geometrical confinement is extreme in the constrictions of the foam network for $\lambda \approx 1$, the concept of effective viscosity makes sense in foam nodes where the suspension is effectively sheared. Moreover, this effect is specific to foams due to the interfacial mobility which allows the particles to flow easily in constrictions [21]. For $\varphi_p \gtrsim 0.2$ Fig. 4 shows deviations in the viscosity corresponding to different values of $\lambda$. Moreover these deviations increase as a function of the particle volume fraction. In fact, the data corresponding to each value of $\lambda$ define a distinct curve and can be fitted with the Krieger-Dougherty relation: $\tilde{\mu}_{eff}(\varphi_p) = \left(1 - \varphi_p/\varphi_p^{crit}\right)^{-2.5\varphi_p^{crit}}$, where $\varphi_p^{crit}$ is the critical particle volume fraction for which the viscosity diverges [22]. As shown in Fig. 4, the critical particle volume fraction obtained by fitting the data depends on $\lambda$: $\varphi_p^{crit} = 0.57, 0.53, 0.50$ and $0.46$ for $\lambda = 0.13, 0.43, 0.65$ and $0.87$ respectively. The physical meaning of $\varphi_p^{crit}$ is usually interpreted as a consequence of the particle packing at $\varphi_p^*$ [23,24,14], i.e. $\varphi_p^{crit} \cong \varphi_p^*$. Therefore, in the following we seek for a physical interpretation for the reported evolution of $\varphi_p^*$ as a function of $\lambda$. In doing so we determine the packing fraction of particles in the structural elements of the foam network, namely the nodes and the Plateau borders, i.e. $\varphi_{p,node}^*$ and $\varphi_{p,Pb}^*$ respectively. First, $\varphi_{p,node}^*$ can be estimated from existing results for bisdisperse packings of spheres [25,26]. Whereas monodisperse assemblies of fine or coarse particles have the same bulk packing fraction ($\varphi_{bulk}^*$), the overall packing fraction of bidisperse assemblies ($\varphi_{bidisperse}^*$) depends on both $x_F$, the volume fraction of fine particles in the mixture and $\Lambda$, the coarse to fine particle size ratio. We are interested in situations characterized by $\Lambda \gg 1$, where the fine particles are sufficiently small to fill the spaces in the packing of coarse particles. In such a case, the maximum overall packing fraction is $\varphi^{max}_{bidisperse} = \varphi^*_{bulk} + (1 - \varphi^*_{bulk})\varphi^*_p(\Lambda)$, where $\varphi^*_{bulk}$ refers to the packing of coarse particles and $\varphi^*_p(\Lambda)$ is the packing fraction for the fine particles confined in the spaces formed by the packed coarse particles. For $\Lambda \rightarrow +\infty$, $\varphi^*_p(\Lambda) = \varphi^*_{bulk}$, but due to wall effects $\varphi^*_p(\Lambda) < \varphi^*_{bulk}$ for any finite value of $\Lambda$. Models accounting for the wall effect in mixtures of spheres have been proposed and here we refer to the model of de Larrard et al. [26,27]. For large $\Lambda$ values $\varphi^*_{bidisperse}$ is given by the following set of equations:

$$\begin{cases} \varphi^*_{bidisperse} = \min(\varphi_C, \varphi_F) \\ \varphi_C = \dfrac{\varphi^*_{bulk}}{1 - x_F} \\ \varphi_F = \dfrac{\varphi^*_{bulk}}{1 - (1 - x_F)[1 - \varphi^*_{bulk} + b_{CF}(\varphi^*_{bulk} - 1)]} \end{cases} \quad (2)$$

where $b_{CF} = [1 - (1 - 1/\Lambda)^{1.79}]^{0.82}$ [27] is the function accounting for the geometrical wall effect. $\varphi^*_{bidisperse}$ is plotted in the inset of Fig. 5 and it shows how $\varphi^{max}_{bidisperse}$ decreases due to wall effects as $\Lambda$ decreases ($\Lambda \propto \lambda^{-1}$). As the geometry of a foam node differs from that resulting from the contacting coarse spheres discussed above, one as to define an equivalent coarse sphere radius for foam. The shape of foam nodes is imposed by capillary forces and the Young-Laplace law implies that the mean curvature is approximately constant for the node surface, i.e. $H = 1/r_{n,1} + 1/r_{n,2} \approx cte$, where $r_{n,1}$ and $r_{n,2}$ are the two principal radii of curvature. At the node ends, where it connects to Plateau borders, these radii can be approximated by $r_{n,1} \gg r_{n,2} \approx r_{Pb}$, so that $H \approx 1/r_{Pb}$. At the centre of the node surface, the two radii take the same value, i.e. $r_{n,1} = r_{n,2} = r_n$ , and the resulting mean curvature writes: $H = 2/r_n$. Therefore, the central area of the node surface can be described by a spherical cap of radius $r_n \approx 2r_{Pb}$, showing how the two principal radii evolve from the node ends to the central area. In order to average this evolution, one can determine the radius of spheres forming a tetrahedral pore which volume is equal to that of a foam node, $v_n \approx 0.9r^3_{Pb}$ as calculated above. The volume of a tetrahedral pore formed by 4 contacting spheres of radius $R_{coarse}$ is given by $v_{tetra} = 0.245R^3_{coarse}$ [5], which provides an equivalent coarse sphere radius for foam nodes: $R_{coarse} \approx 1.5r_{Pb}$. Thus, the relation between $\Lambda$ and $\lambda$ reads $\Lambda = 2R_{coarse}/d_p = 1.5/(2/\sqrt{3} - 1)\lambda$ and we plot $\varphi^*_{p,node} \equiv \varphi^*_p(\lambda) = (\varphi^{max}_{bidisperse} - \varphi^*_{bulk})/(1 - \varphi^*_{bulk})$ in Fig. 5. Note that (i) the particular choice for $\varphi^*_{bulk}$ has no influence on $\varphi^*_{p,node}$ and (ii) the choice made for the ratio $R_{coarse}/r_{Pb}$ has a very limited influence on $\varphi^*_{p,node}(\lambda)$ within the investigated $\lambda$ range. Fig. 5 shows that the decrease of the packing fraction of particles confined in a foam node reaches 25% as $\lambda$ raises up to unity. For practical purposes, the curve $\varphi^*_{p,node}(\lambda)$ can be approximated by the polynomial curve: $\varphi^*_{p,node}(\lambda)/\varphi^*_{bulk} \simeq 1 - \lambda/3 + 0.1\lambda^2$.

Now we turn to the determination of $\varphi^*_{p,Pb}$ for particles confined in Plateau borders. As far as we know, this problem has never been considered in literature, which justifies the experimental and theoretical elements we develop in the following. We perform a simple packing experiment in a straight solid Plateau border: monodisperse glass beads of diameter $d_p$ = 1.5 – 10 mm are poured in the space between 3 vertical PMMA cylinders in contact of radii $r_{Pb}$ = 11.5 – 40 mm. Bead density is measured and then, from the height and the mass of the beads deposited in this Plateau border geometry, we deduce the particle packing fraction as a function of the confinement parameter $\lambda$. In Fig. 6a, the measurements reveal an overall decrease of $\varphi^*_{p,Pb}$ with $\lambda$, illustrating the increasing importance of both wall effect – the local density is lower at the wall than in the bulk – and corner effect – the 3 corners of the Plateau border are not accessible to particles. An analytical expression of this decrease can be derived by taking into account these two effects:

$$\varphi^*_{p,Pb} = \frac{\varphi^*_{wall}S_{wall} + \varphi^*_{bulk}S_{bulk}}{S_{Pb}} \quad (3)$$

where $S_{Pb} = \left(\sqrt{3} - \frac{\pi}{2}\right) r_{Pb}^2$ is the cross-section of the Plateau border and $S_{wall}$ (resp. $S_{bulk}$) is the area covered by beads packed at $\varphi^*_{wall}$ (resp. $\varphi^*_{bulk}$) as shown in Fig. 6b. The wall effect is approached by considering the ordered configuration, i.e. $\frac{\varphi^*_{wall}}{\varphi^*_{bulk}} = \frac{\varphi^*_{plane}}{\varphi^*_{FCC}}$, where $\varphi^*_{plane} = \frac{\pi}{3\sqrt{3}}$ is the maximum volume fraction of a bead monolayer in a triangular lattice between two planes and $\varphi^*_{FCC} = \frac{\pi}{3\sqrt{2}}$ is the volume fraction of a faced centered cubic packing. $S_{wall}$, $S_{bulk}$ and $S_{corner}$ (the corner area that is not accessible to the particles), are derived from simple geometric considerations:

$$\begin{cases} S_{wall} = \left(\frac{\pi}{2} - 3\alpha\right) r_{Pb}^2 \tan^2\alpha + 3\left(\frac{\pi}{2} - \alpha\right) r_p^2 \\ S_{corner} = 3(\tan\alpha - \alpha) r_{Pb}^2 - 3\left(\frac{\pi}{2} - \alpha\right) r_p^2 \\ S_{bulk} = S_{Pb} - S_{wall} - S_{ex} \end{cases} \quad (4)$$

where the angle $\alpha$ is shown in Fig. 6b. From these expressions, Eq. 3 is plotted in Fig. 6a and it is found to describe well the overall decrease measured for $\varphi^*_{p,Pb}$. Eq. 3 can be approximated by $\varphi^*_{p,Pb}(\lambda)/\varphi^*_{bulk} = 1 - 0.7\lambda + 0.08\lambda^2$.

As $\lambda$ > 0.55, the experimental data show large fluctuations, due to ordering induced by increasing confinement effects. Several ordered packings – from 1 sphere ($\lambda$ = 1) to 4 spheres ($\lambda$ = 0.530) in the Plateau border cross-section – have been calculated. They are reported in Fig. 6a and are illustrated in the appendix. One can derive analytically (see the appendix) the transition between two close configurations as illustrated by the lines in Fig. 6a These results increase our knowledge on sphere packings with geometric constraints such as those obtained for the cylindrical channels geometry [28].

$\varphi^*_{p,node}$ and $\varphi^*_{p,Pb}$ are plotted in Fig. 7 against the critical particle volume fractions reported from Fig 4. The experimental data are found to be in good agreement with $\varphi^*_{p,node}$ within the whole range of $\lambda$-values. This good agreement is due to the fact that the studied foams are rather wet, i.e. most of the suspension is confined in the nodes. This suggests that the geometrical approximation based on bidisperse mixtures of spheres is sufficient to describe this confinement in wet foams. Unfortunately our experimental setup does not allow to explore the behavior of dry particulate foams at high $\varphi_p$ , but the relevance of the lower bound, i.e. $\varphi^*_{p,Pb}$, certainly deserves a dedicated study.

## 5. Conclusion

We performed drainage experiments of particulate foams, where a granular suspension is confined within the interstitial pore space of the foam. Under our experimental conditions, the particles are not trapped individually by the constrictions of the network. We observed the jamming transition when the particle volume fraction reaches a critical value $\varphi^*_p$, that is found to be very sensitive to the particle size. $\varphi^*_p$ is unexpectedly low due to confinement effects when the particles pack into the geometrical elements of the foam network. We model quantitatively the overall behavior of the suspension – from flow to jamming – by taking into account explicitly the divergence of its effective viscosity at $\varphi^*_p$. Our complete study of the geometrical confinement suggests that even lower $\varphi^*_p$ values could be reached by using dryer foams, for which the proportion of liquid contained in the Plateau borders is significantly raised. Beyond the scope of drainage, the reported jamming transition is expected to have a deep significance for all aspects related to particulate foams: rheology and ripening of liquid foams, and mechanics of cellular solids.

## Aknowledgements

We thank G. Ovarlez and X. Chateau for stimulating discussions, D. Hautemayou and C. Mézière for technical support. We gratefully acknowledge financial support from Agence Nationale de la Recherche (Grant No. ANR-13-RMNP-0003-01) and French Space Agency (convention CNES/70980).

## Appendix: Sphere packings in a Plateau Border

Analytical expressions for $\varphi^*_{p,Pb}$ are derived from simple geometrical considerations and they are reported in Table 1. These configurations are represented by the star-symbol on Fig. 6a.

**Table 1**: ordered sphere packings in a Plateau border for different confinement ratio illustrated by the star-symbol on Fig. 6a.

| | $\boldsymbol{\lambda}$ | $\boldsymbol{\varphi^*_{p,Pb}}$ |
|---|---|---|
| 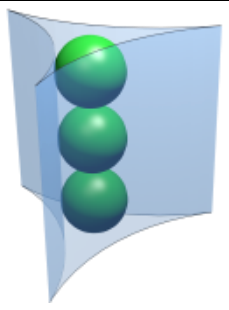 | $\lambda_1 = 1$ | $\varphi^*_1 = \dfrac{2\pi\left(2-\sqrt{3}\right)^2}{9\left(\sqrt{3}-\frac{\pi}{2}\right)} \approx 0.311$ |
| 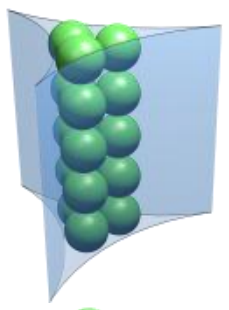 | $\lambda_3 = \dfrac{5\sqrt{3}-6\sqrt{2}}{2-\sqrt{3}} \approx 0.653$ | $\varphi^*_3 = \dfrac{2\pi\left(5-2\sqrt{6}\right)^2}{\sqrt{3}-\frac{\pi}{2}} \approx 0.398$ |
| 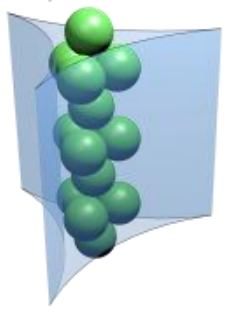 | $\lambda_{3'} = \dfrac{5\sqrt{3}-6\sqrt{2}}{2-\sqrt{3}} \approx 0.653$ | $\varphi^*_{3'} = \dfrac{2\pi\sqrt{6}\left(5-2\sqrt{6}\right)^2}{3\left(\sqrt{3}-\frac{\pi}{2}\right)} \approx 0.325$ |
| 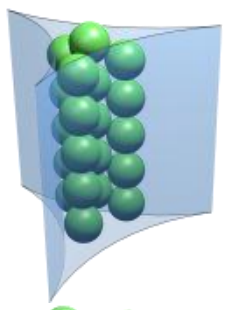 | $\lambda_{3.5} = \dfrac{\sqrt{3}-\sqrt{2.5}}{2-\sqrt{3}} \approx 0.563$ | $\varphi^*_{3.5} = \dfrac{4\pi\left(\sqrt{6}-\sqrt{5}\right)^2}{9\left(\sqrt{3}-\frac{\pi}{2}\right)} \approx 0.394$ |
| 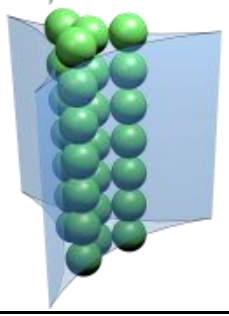 | $\lambda_4 = \dfrac{2+\sqrt{3}-2\sqrt{1+\sqrt{3}}}{3\left(2-\sqrt{3}\right)} \approx 0.530$ | $\varphi^*_4 = \dfrac{8\pi\left(2-\sqrt{3}\right)^2}{9\left(\sqrt{3}-\frac{\pi}{2}\right)}\lambda^2 \approx 0.350$ |

From these particular patterns, we derive the transitions $\varphi^*_{1\to3}$, from 1 particle to 3 particles in a Plateau border cross-section, $\varphi^*_{3'\to3.5}$ , from 3 particles to 3.5 particles and $\varphi^*_{3.5\to4}$, from 3.5 to 4 particles ($d_p = 2r_p$):

$$\frac{l}{r_p} = \left(2+\sqrt{3}\right)\lambda^{-1} - \sqrt{1+\left(4\sqrt{3}+6\right)\lambda^{-1}}$$

$$\begin{cases} \dfrac{z_1}{r_p} = \sqrt{4-3\left(\dfrac{l}{r_p}\right)^2} \\ \varphi^*_{1\to3} = \dfrac{4\pi\left(7-4\sqrt{3}\right)}{9\left(\sqrt{3}-\frac{\pi}{2}\right)} \dfrac{\lambda^2}{\max\left(\frac{z_1}{r_p};\dfrac{2}{3}\right)} \end{cases}$$

$$\begin{cases} \dfrac{z_3}{r_p} = \sqrt{4 - \left(\dfrac{l}{r_p}\right)^2} \\ \varphi^*_{3'\to 3.5} = \dfrac{8\pi\left(7 - 4\sqrt{3}\right)}{9\left(\sqrt{3} - \frac{\pi}{2}\right)} \dfrac{\lambda^2}{\frac{z_3}{r_p}} \end{cases}$$

$$\varphi^*_{3.5\to 4} = \frac{8\pi\left(7 - 4\sqrt{3}\right)}{9\left(\sqrt{3} - \frac{\pi}{2}\right)} \lambda^2$$

These expressions are illustrated by the fine lines on Fig. 6.

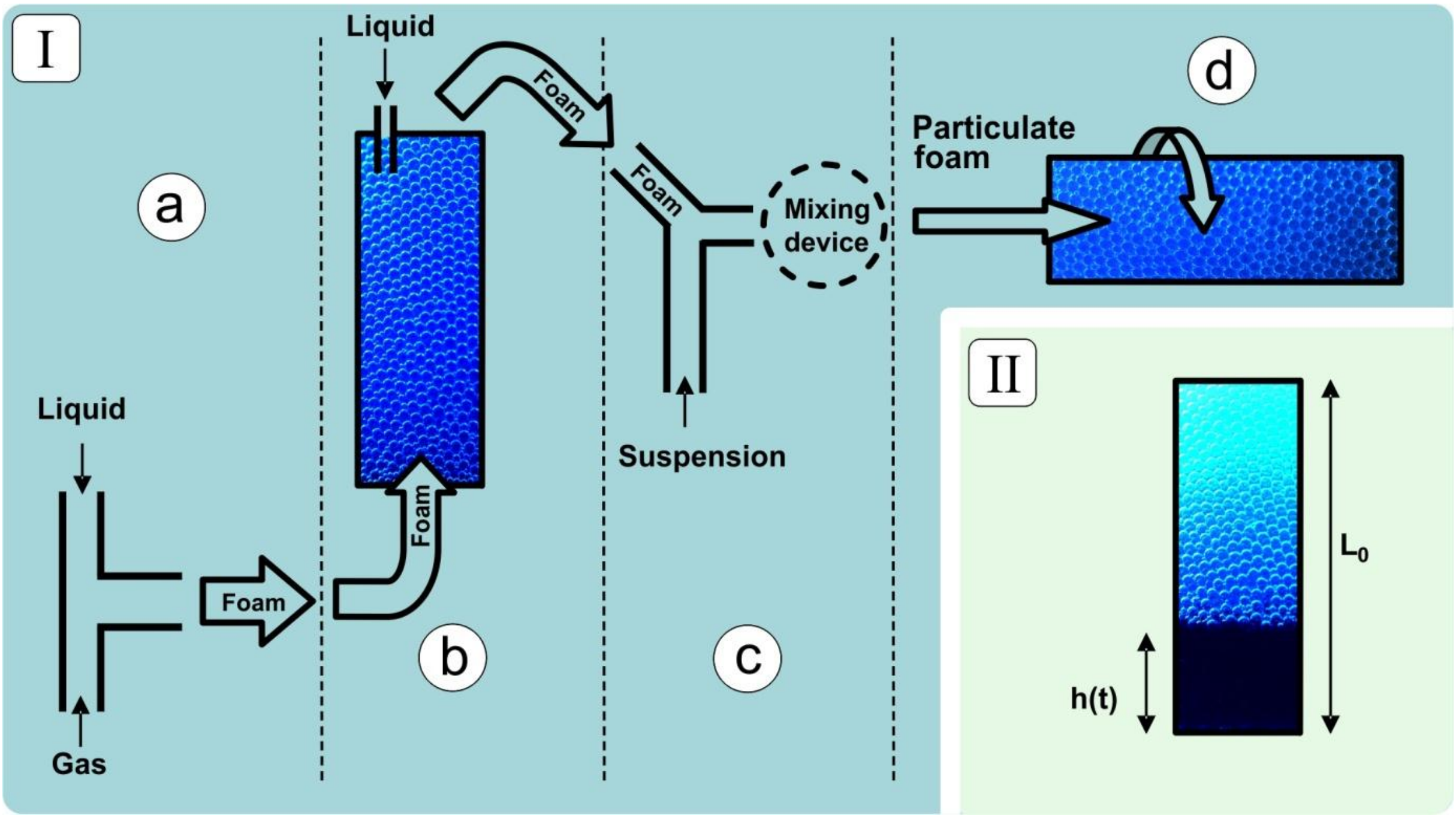

Fig. 1: Experimental setup. I- Production of particulate foams: monodisperse bubbles are generated from the simultaneous injection of gas and foaming solution through a T-junction (a). The bubbles are released at the bottom of a column partially filled with the foaming solution and foam is produced. Imbibition with the same foaming solution allows setting the gas fraction over the whole foam sample (b). Once the foam has filled the column, it is injected along with a granular suspension in a small device in order to obtain the final mixture (c), the proportion of each phase being accurately controlled during this stage. The mixture is continuously introduced in a horizontal column where rotation allows for gravity effects to be compensated (d). II- Study of drainage: after the generation step, the rotating motion is stopped and the column is turned to the vertical. A camera is then used to follow the evolution for the position of the foam/liquid transition, from which the drainage velocity is determined.

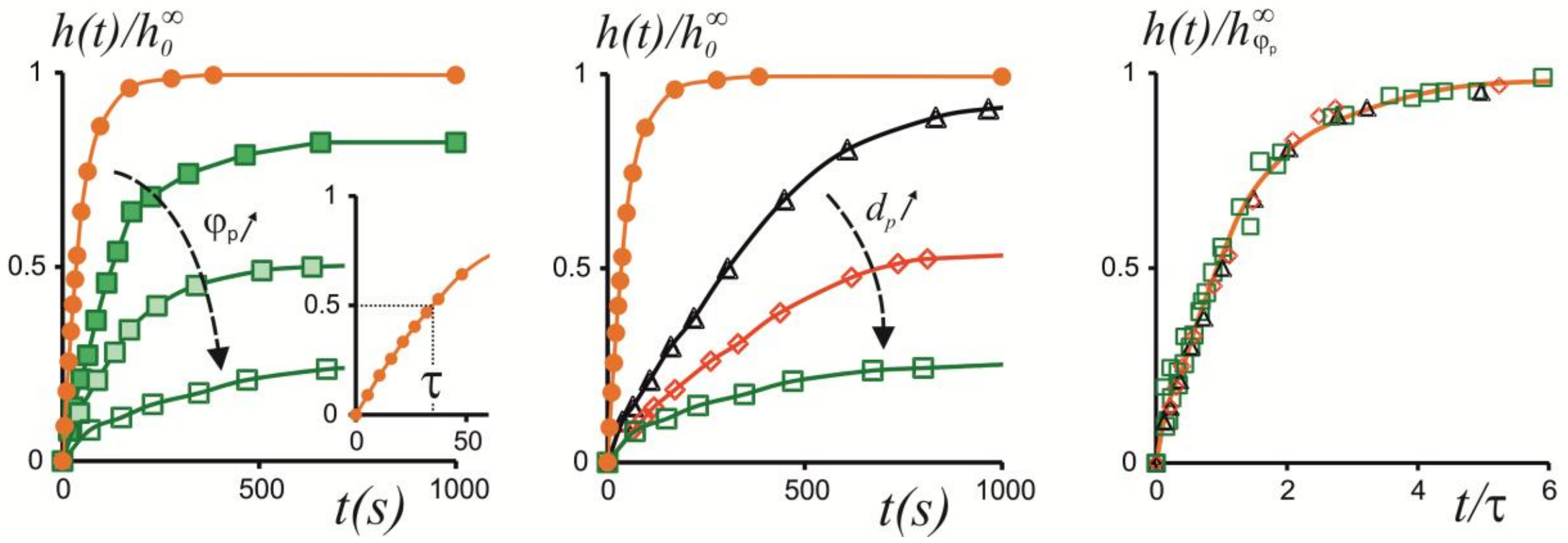

(a) (b) (c)

Fig. 2: Temporal evolution of the reduced height of liquid/suspension drained out of the foam. (a) Effect of the particle volume fraction at fixed particle size $d_p = 40$ µm: $\varphi_p = 0$ (●), 0.16 (■), 0.37 (■) and 0.45 (□); inset: zoom on the linear regime of the particle-free foam. (b) Effect of the particle size at fixed particle volume fraction $\varphi_p = 0.45$: $d_p = 6$ µm (Δ), 20 µm (◇) and 40 µm (□). (c) Rescaled drainage curves from all the data of (a) and (b).

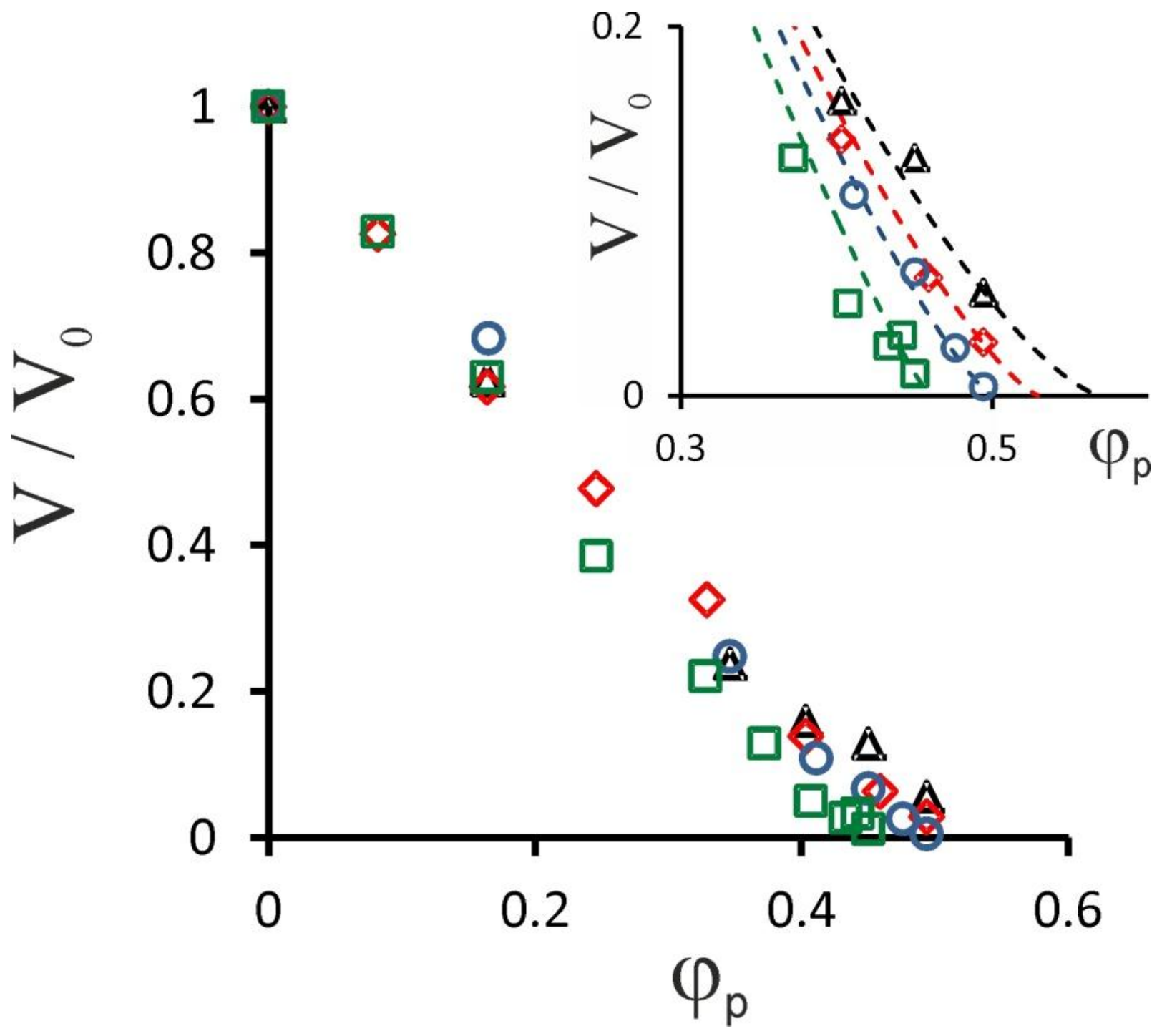

Fig. 3: Reduced drainage velocity as a function of the particle volume fraction for several particle sizes : $d_p = 6$ μm (△), 20 μm (◇), 30 μm (○), and 40 μm (□). Inset: zoom on vanishing drainage velocities.

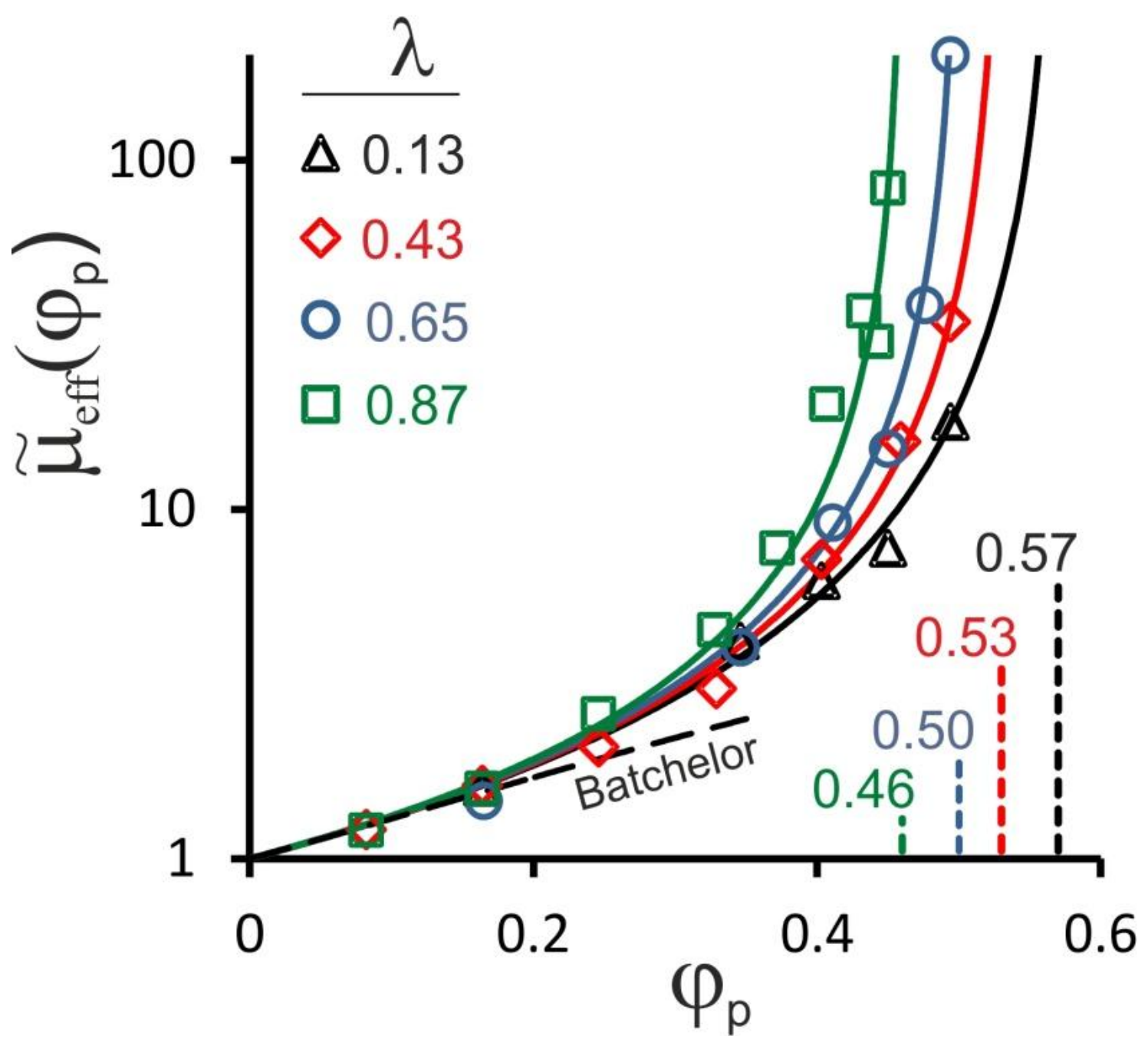

Fig. 4: Reduced effective viscosity of the suspension as a function of the particle volume fraction for several $\lambda$ values. The solid lines correspond to Krieger-Dougherty curves using the critical particle volume fractions reported on the abscissa for each $\lambda$ value. The dash line corresponds to $\tilde{\mu}_{eff}(\varphi_p) = 1 + 2.5\varphi_p + 6.95\varphi_p^2$.

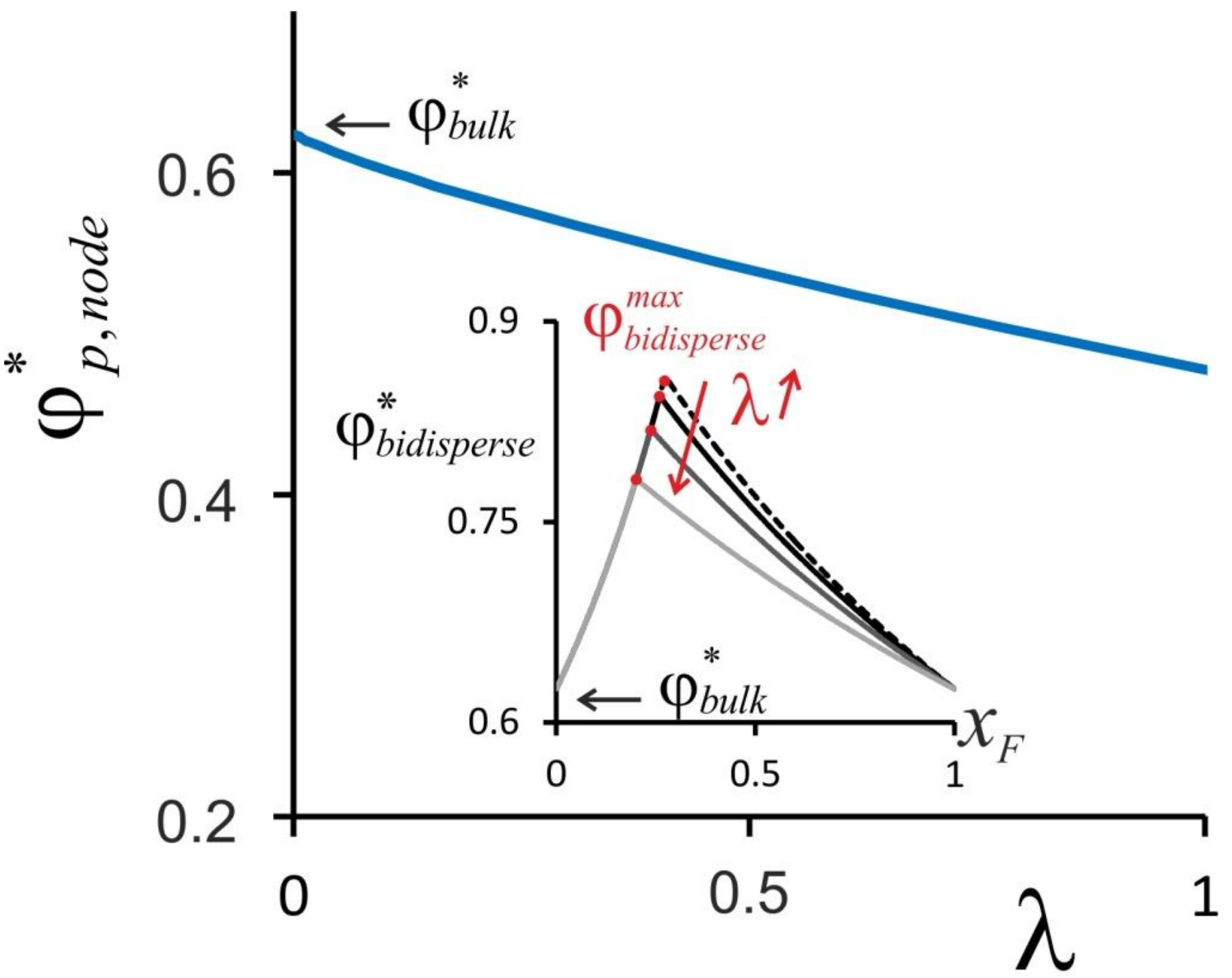

Fig. 5: Packing fraction of spheres confined in a foam node ($\varphi^*_{bulk}$ is the packing fraction within unconfined conditions). Inset: Packing fraction of bidisperse assemblies of coarse and fine particles as a function of the proportion of fine particles – computed from eq. (2). The maximum value is shown to decrease as the coarse to fine size ratio decreases from ∞ to 6.5, or equivalently as $\lambda$ increases from 0 to 1.

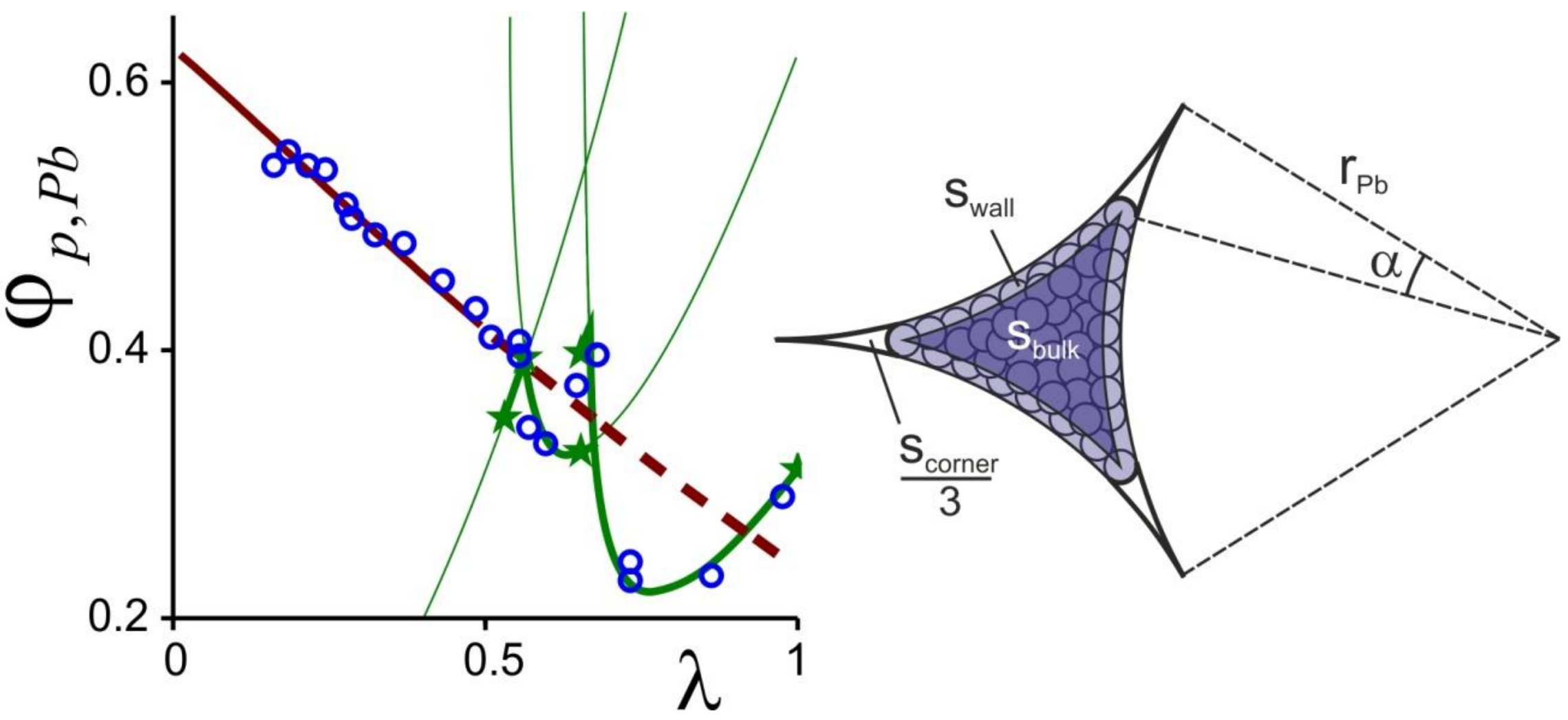

(a) (b)

Fig. 6. Packing of spheres in an ideal Plateau border. (a) (○) experimental data for glass beads poured in the space between 3 cylinders in contact, thick brown line: Eq. 3, thin green lines: Eqs. 5, 6 and 7 (appendix) and (★) ordered sphere packings (appendix) (b) Sketch of the Plateau border cross-section filled with particles.

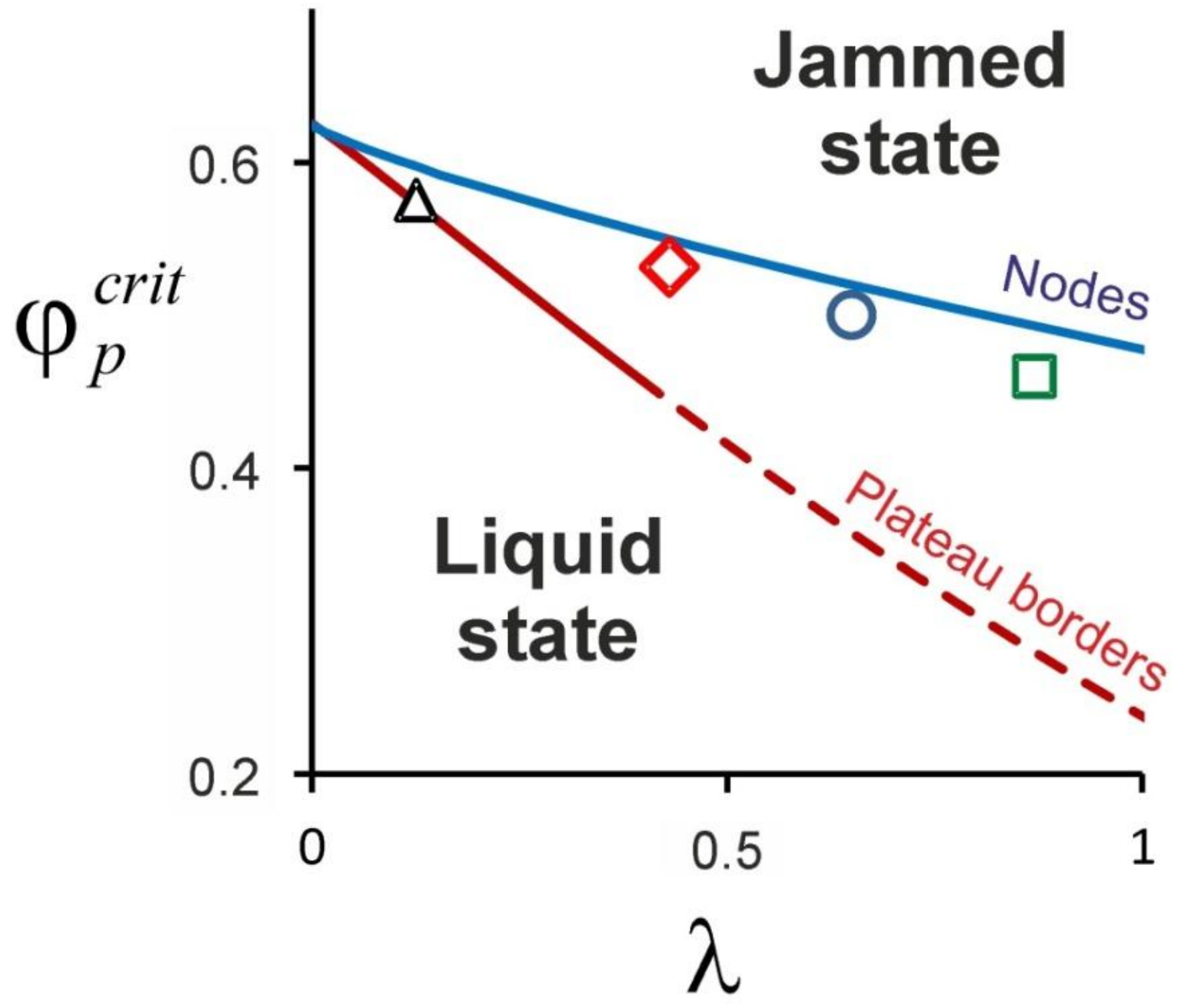

Fig. 7: Critical volume fractions measured for particles suspensions confined in foams (the symbols are the same than those presented in Fig. 4). The lines correspond to packing fractions calculated for nodes and Plateau borders (respectively $\varphi^{*}_{p,node}(\lambda)$ and $\varphi^{*}_{p,Pb}(\lambda)$ defined in the text).